\documentclass{PoS}

\title{Electric and magnetic Landau-gauge gluon propagators in 
       finite-temperature SU(2) gauge theory}

\ShortTitle{Electric and magnetic gluon propagators}

\author{Attilio Cucchieri$^{ab}$, \speaker{Tereza Mendes}$^a$\\
\llap{$^a$}Instituto de F\'\i sica de S\~ao Carlos, Universidade de S\~ao Paulo, \\
           Caixa Postal 369, 13560-970 S\~ao Carlos, SP, Brazil\\
\llap{$^b$}Ghent University, Department of Physics and Astronomy, \\
           Krijgslaan 281-S9, 9000 Gent, Belgium \\
E-mail: \email{attilio@ifsc.usp.br}, \email{mendes@ifsc.usp.br}}

\abstract{We perform lattice simulations in pure-SU(2) Yang-Mills theory 
to investigate how the infrared behavior of electric and magnetic gluon 
propagators in Landau gauge is affected by temperature. 
We consider the largest lattices to date, in an attempt to 
keep systematic errors under control. Electric and magnetic 
screening masses are calculated through an Ansatz from the 
zero-temperature case, based on complex-conjugate poles for the 
momentum-space propagators. As recently reported in \cite{Cucchieri:2011ga}, 
we find good fits to the proposed form at all temperatures considered, 
with different ratios of real to imaginary part of the pole masses for
the longitudinal (electric) and transverse (magnetic) propagators.
The behavior of the magnetic propagator $D_T(p)$ is in agreement with 
the dimensional-reduction picture, showing infrared suppression (with a 
turnover in momentum) and violation of spectral positivity at all 
nonzero temperatures considered. The longitudinal propagator $D_L(p)$
appears to reach a plateau at small momenta and is subject to severe 
finite-$N_t$ effects around the critical temperature $T_c$. As a 
consequence, only lattices with temporal extent $N_t > 8$ seem to be 
free from systematic errors. 
After these errors are removed, the infrared-plateau value is considerably 
reduced around the transition and the sharp peak observed previously for
this quantity at $T_c$ is no longer present. 
The resulting infrared behavior for $D_L(p)$ at $T_c$ is essentially the 
same as for $0.5 T_c$. An investigation of the temperature range between 
$0.5 T_c$ and $T_c$ reveals that a less pronounced (finite) peak may 
occur at smaller temperatures, e.g.\ $T\approx 0.9 T_c$.}

\FullConference{The many faces of QCD\\
		November 2-5, 2010\\
		Gent Belgium}

\def\spose#1{\hbox to 0pt{#1\hss}}
\def\ltapprox{\mathrel{\spose{\lower 3pt\hbox{$\mathchar"218$}}
 \raise 2.0pt\hbox{$\mathchar"13C$}}}
\def\gtapprox{\mathrel{\spose{\lower 3pt\hbox{$\mathchar"218$}}
 \raise 2.0pt\hbox{$\mathchar"13E$}}}

\begin{document}


\section{Introduction}

At zero temperature, Landau-gauge gluon and ghost propagators are 
believed to be closely related to the confinement mechanism in 
Yang-Mills theories, following the so-called Gribov-Zwanziger 
scenario \cite{Zwanziger:1991gz,Zwanziger:2011yy}.
A key feature of this scenario, formulated for momentum-space 
propagators, is that the gluon propagator $D(p)$ should be
{\em suppressed} in the infrared limit. Such a suppression
is associated with violation of spectral positivity, which is 
commonly regarded as an indication of gluon confinement. 
Lattice studies (see \cite{Cucchieri:2010xr} for a review) have 
confirmed the suppression of $D(p)$ in the infrared limit
and have also observed violation of reflection positivity for 
the real-space gluon propagator \cite{Cucchieri:2004mf}.
The infrared data for $D(p)$ are well fitted by a Gribov-Stingl form
(see e.g.\ \cite{Cucchieri:2003di}), which generalizes the form 
originally proposed by Gribov, based on a propagator with a pair of 
complex-conjugate poles. These poles can be associated with complex 
values for dynamically generated masses, a behavior in agreement with 
the massive (or decoupling) solution of Schwinger-Dyson equations 
\cite{Aguilar:2008xm}. The same behavior is obtained in the
refined Gribov-Zwanziger framework \cite{Dudal:2008sp,Dudal:2008rm}.
(See also \cite{Kondo:2011ab} for a very recent proposal in maximally
Abelian gauge.)

At high temperatures, on the other hand, one expects to observe 
Debye screening of the color charge, signaled by screening masses/lengths 
that can in principle be obtained from the gluon propagator \cite{Gross:1980br}.
More specifically, chromoelectric (resp.\ chromomagnetic) screening 
will be related to the longitudinal (resp.\ transverse) gluon 
propagator computed at momenta with null temporal component, 
i.e.\ with $p_0 = 0$ (soft modes).
In particular, we expect the real-space longitudinal
propagator to fall off exponentially at long distances,
defining a (real) electric screening mass,
which can be calculated perturbatively to leading order.
Also, according to the 3d adjoint-Higgs picture for dimensional 
reduction, we expect the transverse propagator to show a confining 
behavior at finite temperature, in association with a nontrivial 
magnetic mass (see e.g.\ \cite{Cucchieri:2001tw}).
We note that these propagators are gauge-dependent quantities,
and the (perturbative) prediction that the propagator poles should
be gauge-independent must be checked, by considering different gauges.

Although the nonzero-$T$ behavior described above has been verified for 
various gauges and established at high temperatures down to around 
twice the critical temperature $T_c$ \cite{Cucchieri:2001tw,Heller:1997nqa}, 
it is not clear how a screening mass would show up around $T_c$.
In the following, we try to use the knowledge gained in the study of
the zero-temperature case to define temperature-dependent masses
for the region around and below the critical temperature.
We review briefly the existing lattice results for this temperature
range, present our fitting form for the infrared region, show our
preliminary results for gluon propagators on large lattices for 
several values of the temperature and draw our conclusions. 
A more detailed analysis and additional data will be presented shortly
elsewhere \cite{inprep}.

\section{Gluon propagators around and below $T_c$}

The behavior of Landau-gauge gluon and ghost propagators around the
critical temperature $T_c$ has been investigated in \cite{Cucchieri:2007ta}.
That study showed a stronger infrared suppression for the transverse 
propagator $D_T(p)$ than for the longitudinal one $D_L(p)$, confirming 
the dimensional-reduction picture also at smaller temperatures.
[We note here that a recent study \cite{Bornyakov:2010nc}
discusses whether this suppression is consistent with $D_T(0)=0$
and investigates Gribov-copy effects for the propagators.]
It was also found that the ghost propagator is insensitive to the 
temperature.
For the longitudinal gluon propagator, a very interesting
behavior was seen: the data approach a plateau (as a function of the
momentum) in the infrared region and, as a function of temperature,
this plateau shows a sharp peak around the critical temperature. 
The exact behavior
around $T_c$ (e.g.\ whether the peak turns into a divergence at infinite
volume) could not be determined, since relatively small lattices 
were used. All studies mentioned so far are for SU(2) gauge 
theory. 
The momentum-space expressions for the transverse and longitudinal 
gluon propagators $D_T(p)$ and $D_L(p)$ can be found e.g.\ 
in \cite{Cucchieri:2007ta}.

More recently, in \cite{Fischer:2010fx}, further simulations around $T_c$ 
confirmed the above results, and lattice data for the gluon propagator
were used to construct an order parameter for
the chiral/deconfinement transition. More precisely, the authors
use a much finer resolution around $T_c$ and consider the SU(2) and
SU(3) cases. A check of their calculation is done for
the electric screening mass, taken as $D_L(0)^{-1/2}$ and
extracted from the data, where only the $p=0$ raw data point is used.
The considered lattice sizes are still moderate.
[We also mention a very recent study of the $SU(3)$ case, presented
in \cite{Bornyakov:2011jm}.]

Of course, even if an exponential fit of the (real-space) longitudinal gluon
propagator works at high temperature, implying that $D_L(0)^{-1/2}$ is
proportional to the electric screening mass in this limit, it is not 
obvious that this should hold at $T\gtapprox T_c$.
One should therefore consider more general fits.
At $\,T=0$, the momentum-space propagator is well fitted by a
Gribov-Stingl form (see e.g.\ \cite{Cucchieri:2003di}), allowing 
for complex-conjugate poles
\begin{equation}
D_{L,T}(p) \;=\; 
C\,\frac{1\,+\,d\,p^{2 \eta}}
{(p^2 + a)^2 \,+\, b^2}\,.
\label{GSform}
\end{equation}
This expression corresponds to two poles, at
masses $\,m^2 \;=\; a\,\pm\, i b$, where $m \;=\; m_R \,+\, i m_I$.
The mass $m$ thus depends only on $a$, $b$ and not on the normalization $C$.
The parameter $\eta$ should be 1 if the fitting form also describes
the large-momenta region (from our infrared data we get $\eta\neq 1$).
For consistency with the usual definition of electric screening mass,
we expect to observe $\,m_I\to 0\,$ ($\,b\to 0\,$) 
for the longitudinal gluon propagator at high temperature.
Clearly, if the propagator has the above form, then
the screening mass defined by $\,D_L(0)^{-1/2} \,=\, \sqrt{(a^2+b^2)/C}\;\,$
mixes the complex and imaginary masses $\,m_R$ and $m_I\,$ and depends 
on the (a priori arbitrary) normalization $C$.


\section{Results}

We have considered the pure SU(2) case, with a standard Wilson action.
For our runs we employ a cold start, performing a projection on positive 
Polyakov loop configurations. 
Also, gauge fixing is done using stochastic overrelaxation and
the gluon dressing functions are normalized to 1 at 2 GeV. 
We take $\beta$ values in the scaling region
and lattice sizes ranging from $N_s =$ 48 to 192 and from $N_t =$ 2 to 16 
lattice points, respectively along the spatial and along the temporal 
directions.

We note that we have improved our procedure for determining the
physical temperature $T$. Instead of using the value of the lattice 
spacing $a$ in physical units to obtain $T = 1/N_t\, a$, 
we evaluate the ratio $T/T_c$ for a given pair ($\beta$, $N_t$)
by expressing $T$ in terms of the lattice string tension $\sigma$.
More precisely, we consider the ratio $T/T_c = \sqrt{\sigma_c}/\sqrt{\sigma}$, 
where $\sigma$ is the (lattice) string tension evaluated at ($\beta$, $N_t$)
and $\sqrt{\sigma_c}$ is evaluated at the critical coupling $\beta_c$
for the same $N_t$. In this way, one avoids the
inconsistency of obtaining different values for the physical critical 
temperature $T_c$ for different $N_t$'s.
We have taken the $\beta_c$ values for the various $N_t$'s from 
\cite{Fingberg:1992ju}. The string tension has been evaluated using 
the fit given in \cite{Bloch:2003sk}.
This leads to slightly different values of the ratio 
$T/T_c$ as compared with \cite{Cucchieri:2011ga,Cucchieri:2007ta}.

All our data have been fitted to a Gribov-Stingl behavior, as 
described in the previous section (see Eq.\ \ref{GSform}). These
fits are shown here in all plots of $D_{L,T}(p)$, whereas a detailed 
discussion of the associated masses $m_R$, $m_I$ will be presented elsewhere 
\cite{inprep}. We generally find good fits to the Gribov-Stingl 
form (including the full range of momenta), with 
nonzero real and imaginary parts of the pole masses in all cases.
For the transverse propagator $D_T(p)$, the masses $m_R$ and $m_I$ 
are of comparable size (around 0.6 and 0.4 GeV respectively). 
The same holds for $D_L(p)$, but in this case the relative size of 
the imaginary mass seems to decrease with increasing temperature. 

\begin{figure}
\hspace*{-1.5cm}
\includegraphics[height=7.2truecm]{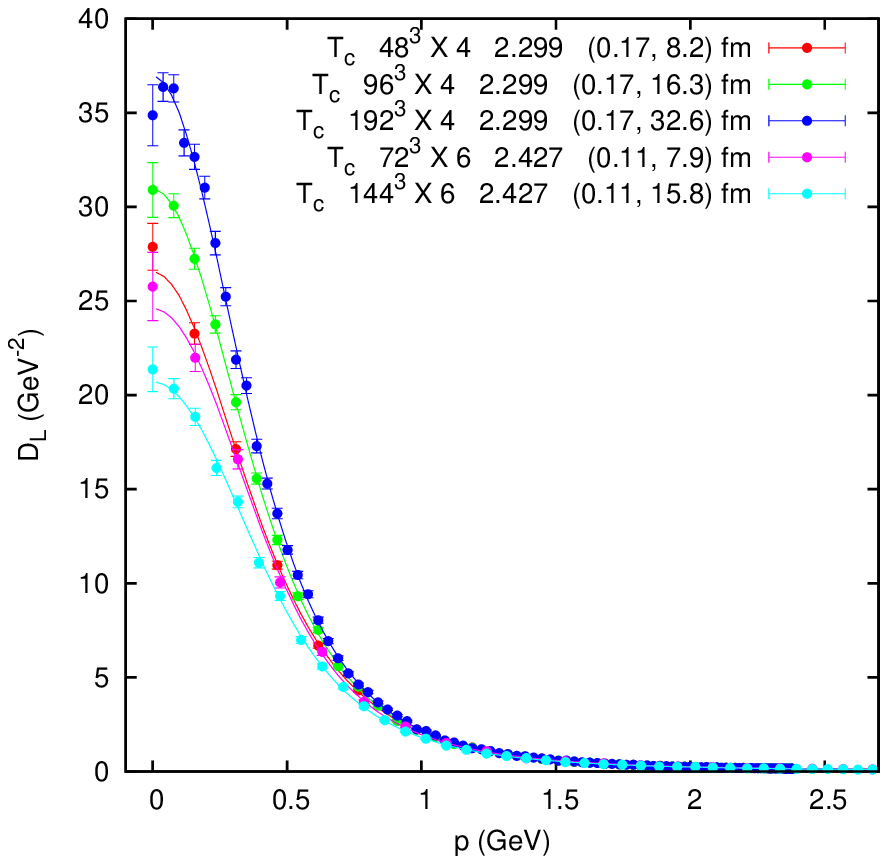}
\hspace*{-2.7cm}
\includegraphics[height=7.2truecm]{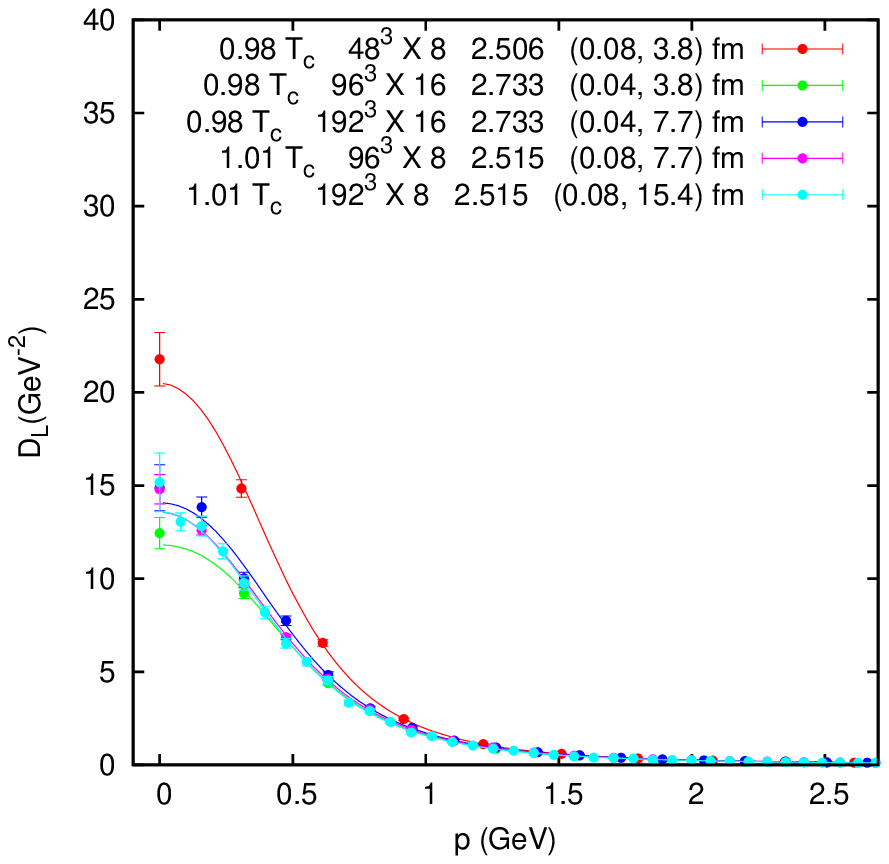}
\caption{Longitudinal gluon propagator at and around
$T_c$, for various lattice sizes and values of $\beta$.
Values for the temperature, $\,N_s^3 \times N_t$, $\,\beta$, lattice 
spacing $a$ and spatial lattice size $L$ (both in fm, in parentheses) 
are given in the plot labels. 
Lattices with $N_t < 8$ and with $N_t \geq 8$ are shown
respectively in the left and right panel.}
\label{DLatTc}
\end{figure}

Our runs were initially planned under the assumption that a temporal
extent $N_t = 4$ might be sufficient to observe the
infrared behavior of the propagators. (Our goal was, then, to increase
$N_s$ significantly, to check for finite-size effects.) For this value 
of $N_t$, the chosen $\beta$ values:
2.2615, 2.2872, 2.299, 2.3045, 2.313, 2.333, 2.5058
yield temperatures respectively of
0.92, 0.96, 1.00, 1.02, 1.05, 1.12, 1.98
times the critical temperature $T_c$.
(See comment above about slight differences in values of
$T/T_c$ for a given $\beta$ in our newest analysis when compared with 
\cite{Cucchieri:2011ga,Cucchieri:2007ta}.)
As seen in Fig.\ \ref{DLatTc}, the assumption that
$N_t = 4$ might be enough is {\em not} verified for the longitudinal 
propagator around the critical temperature, especially in the case 
of larger $N_s$.
Indeed, as $N_s$ is doubled from 48 to 96 and then to 192, we see 
in Fig.\ \ref{DLatTc} (left) that
the infrared value of $D_L(p)$ changes drastically, resulting in a 
qualitatively different curve at $N_s=192$, apparently with a 
turnover in momentum. (Also, in this case the real-space longitudinal 
propagator manifestly violates reflection positivity.)
We took this as an indication that our choice of $N_t=4$ was not 
valid and therefore considered larger values of $N_t$. 
Note that, due to the improved method for introducing physical
units, the data for $N_t < 8$ (left panel of Fig.\ \ref{DLatTc})
lie exactly at $T_c$, while the data at $N_t \geq 8$ (right panel) are
at 0.98 $T_c$ and 1.01 $T_c$. 

As can be seen in Fig.\ \ref{DLatTc} (right), we obtain in this 
way a different picture for the critical behavior of $D_L(p)$. 
Once we use values of $N_t$ that are large enough --- i.e.\ 
$N_t > 8$ for $T\ltapprox T_c$ and $N_t \geq 8$ for $T\gtapprox T_c$ --- 
the curve stabilizes within statistical errors for four different 
combinations of parameters. 
In particular, this includes the two curves at fixed physical volume 
(the blue and the magenta curves).
It is interesting to note that the $N_s$ effects at $T_c$
are significant for $N_t=6$ (with opposite sign with respect to the $N_t=4$ 
case) and are still present for $N_t = 8$ (and maybe also for $N_t = 16$)
slightly below $T_c$, but not immediately above $T_c$. 
Note also that the curves corresponding to the smallest physical 
spatial size (i.e.\ around 4 fm), may show mild finite-physical-size effects. 

\begin{figure}
\hspace*{-1.5cm}
\includegraphics[height=7.2truecm]{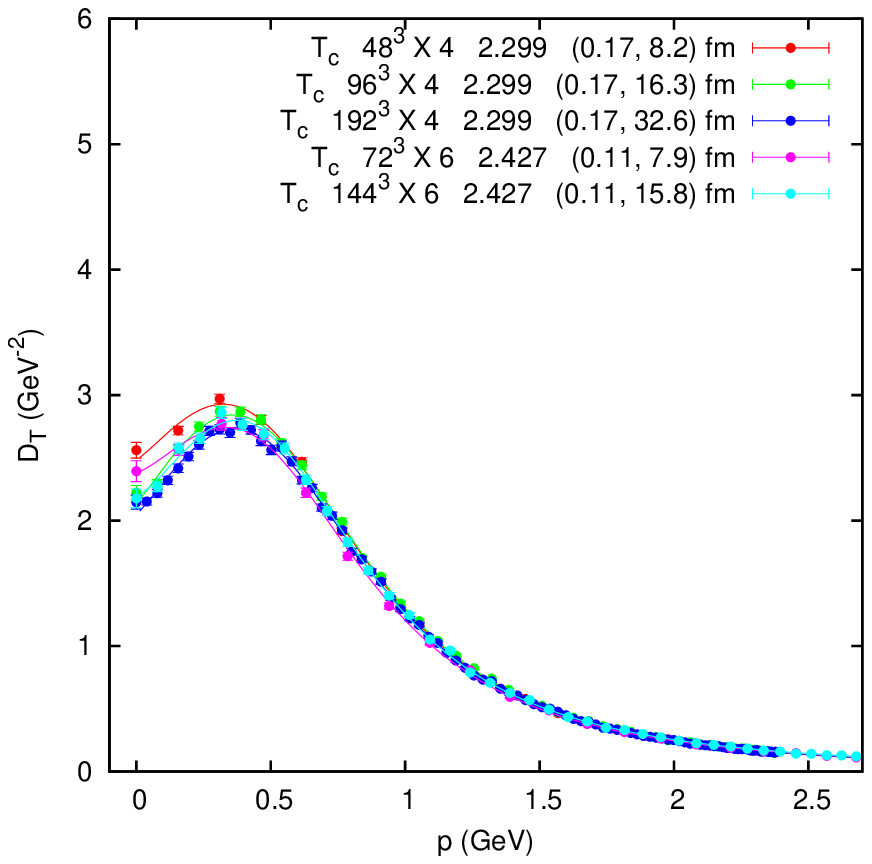}
\hspace*{-2.7cm}
\includegraphics[height=7.2truecm]{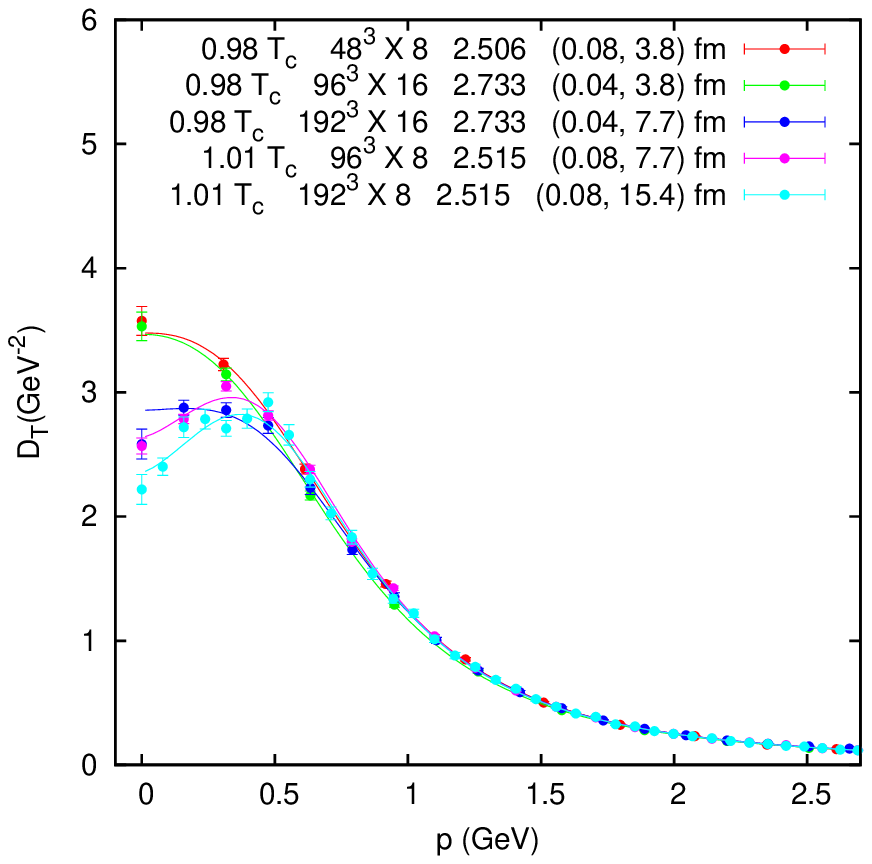}
\caption{Transverse gluon propagator at and around
$T_c$, for various lattice sizes and values of $\beta$.
Values for the temperature, $\,N_s^3 \times N_t$, $\,\beta$, lattice 
spacing $a$ and spatial lattice size $L$ (both in fm, in parentheses) 
are given in the plot labels. 
Lattices with $N_t < 8$ and with $N_t \geq 8$ are shown
respectively in the left and right panel.}
\label{DTatTc}
\end{figure}

In Fig.\ \ref{DTatTc} we show our data for the transverse propagator 
$D_T(p)$ at the critical temperature. In this case, the 
finite-physical-size effects are more pronounced and, in particular,
the lattices with the smallest physical spatial size (the red and green
curves on the right) show qualitatively different behavior when compared 
to the other curves.
On the other hand, $D_T(p)$ does not seem to suffer from the 
same small-$N_t$ effects as $D_L(p)$. Also, we see clearly
the strong infrared suppression of the propagator, with a turnover at
around 400 MeV.


In summary, the transverse propagator $D_T(p)$ shows significant
finite-physical-size effects at $T_c$, while the longitudinal propagator 
$D_L(p)$ is subject to two sources of systematic errors for small $N_t$:
``pure'' small-$N_t$ effects (associated with discretization errors) 
and strong dependence on the spatial lattice size $N_s$ at fixed $N_t$, 
when this value of $N_t$ is smaller than 16.
The latter effect was observed only at $T \ltapprox T_c$, whereas the
former is present in a wider region around $T_c$.
For all investigated values of the temperature, $\,D_L(p)$ seems to 
reach a plateau at small $p$.

\begin{figure}
\hspace*{-1.5cm}
\includegraphics[height=7.2truecm]{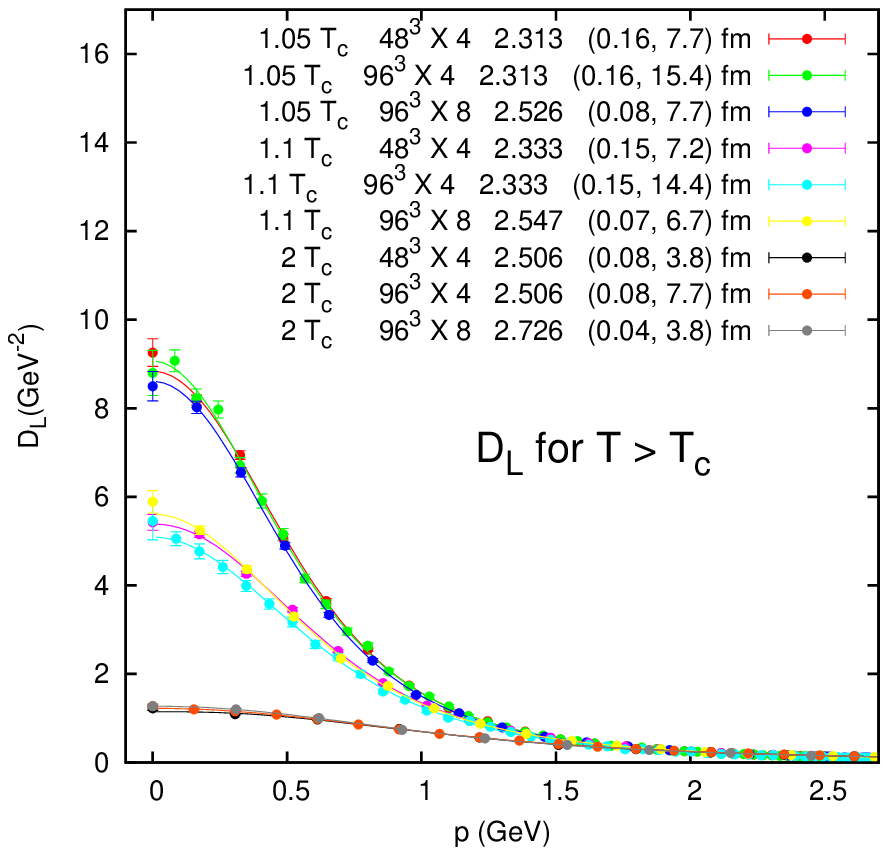}
\hspace*{-2.7cm}
\includegraphics[height=7.2truecm]{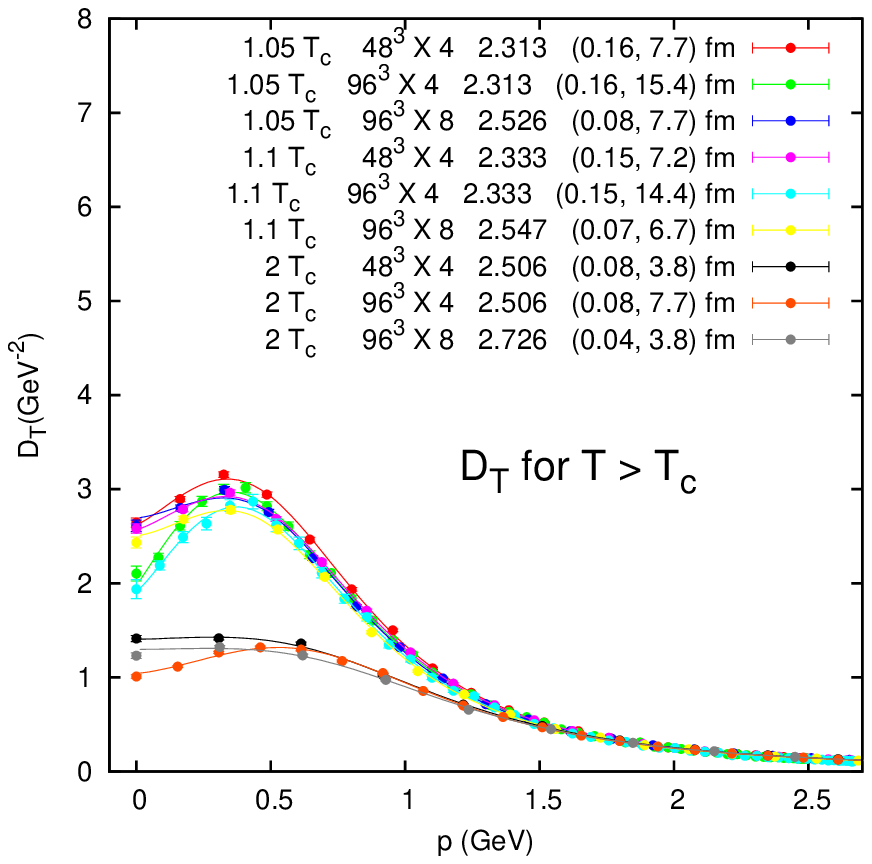}
\caption{Longitudinal (left) and transverse (right) gluon propagator
for $T>T_c$.
Values for the temperature, $\,N_s^3 \times N_t$, $\,\beta$, lattice 
spacing $a$ and spatial lattice size $L$ (both in fm, in parentheses) 
are given in the plot labels.}
\label{DLTaboveTc}
\end{figure}

The plateau value drops significantly for $T\gtapprox T_c$
and then shows a steady decrease. The behavior of $D_L(p)$ and $D_T(p)$ 
for temperatures above $T_c$ is shown in Fig.\ \ref{DLTaboveTc}.
We see, again, that $D_T(p)$ shows finite-physical-size effects
for the smaller lattices (especially for spatial sizes below 4 fm,
but also for 8 fm when compared with 15 fm).
In the $D_L(p)$ data, on the contrary, there are no visible systematic 
effects for the considered lattices at these values of the temperature.

\begin{figure}
\hspace*{-1.5cm}
\includegraphics[height=7.2truecm]{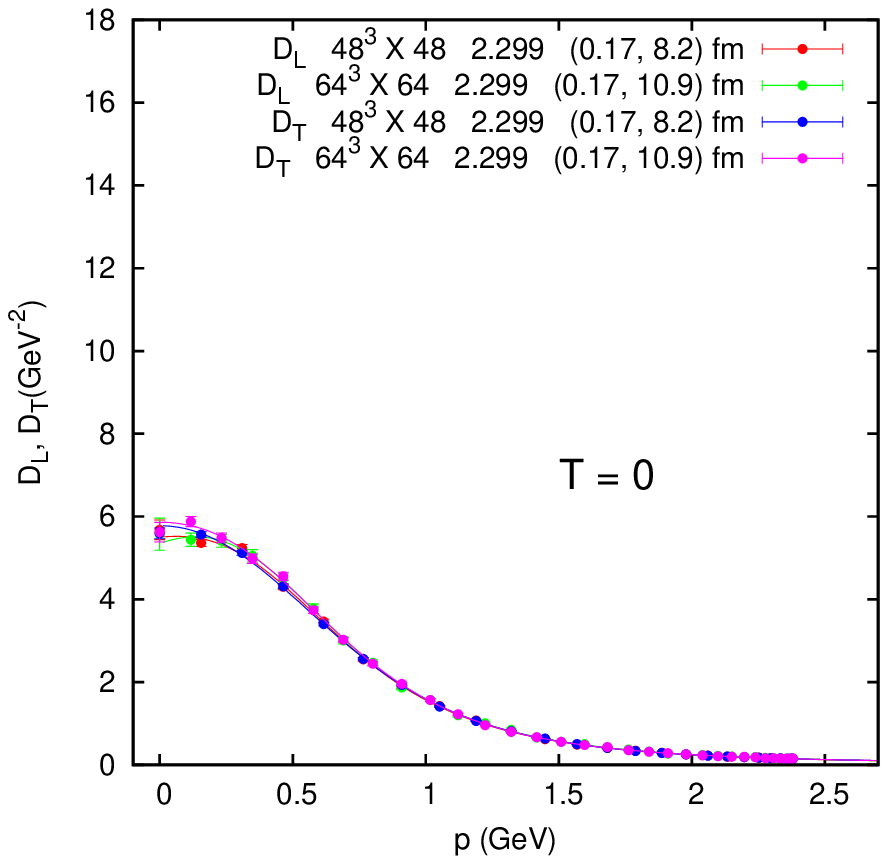}
\hspace*{-2.7cm}
\includegraphics[height=7.2truecm]{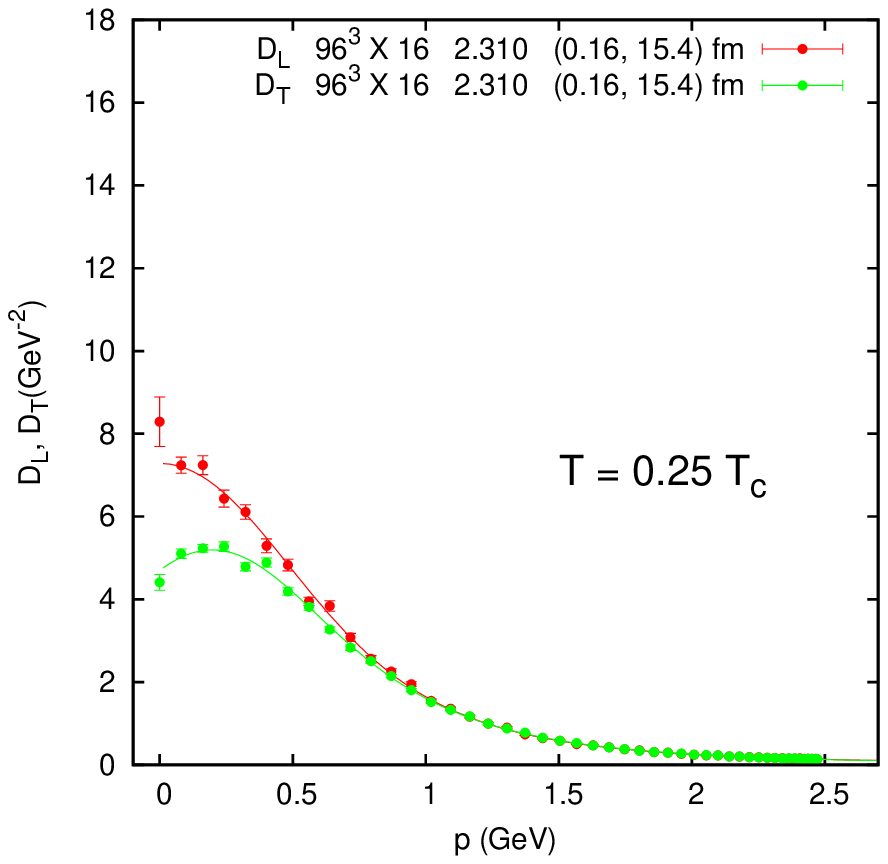}
\vspace*{2mm}
\hspace*{-1.5cm}
\includegraphics[height=7.2truecm]{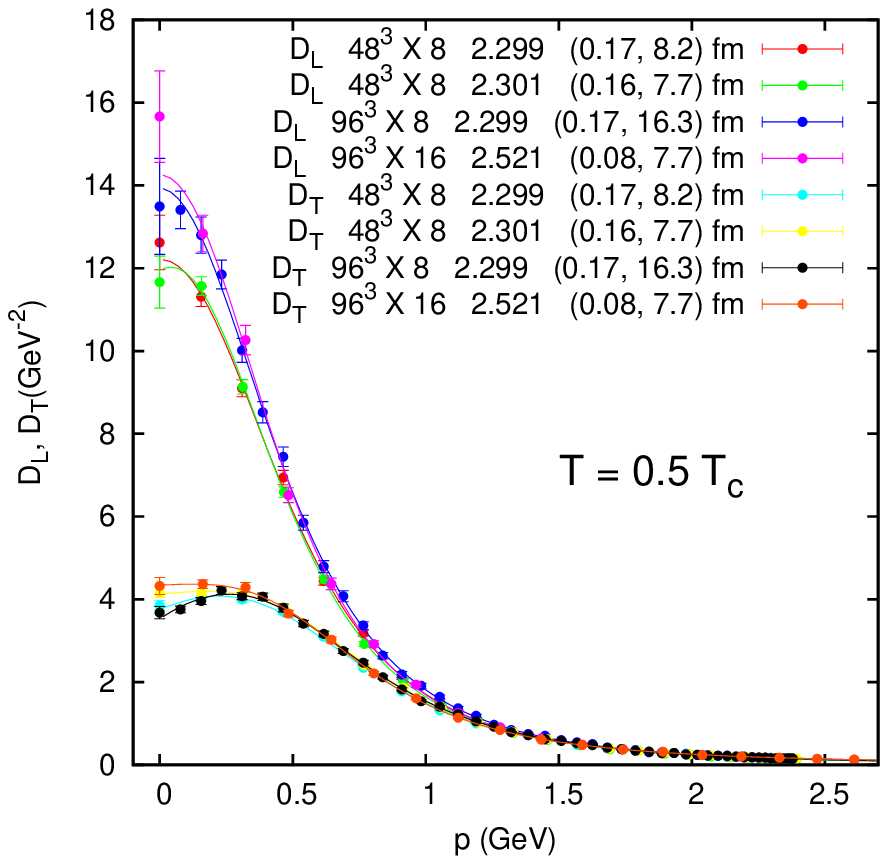}
\hspace*{-2.7cm}
\includegraphics[height=7.2truecm]{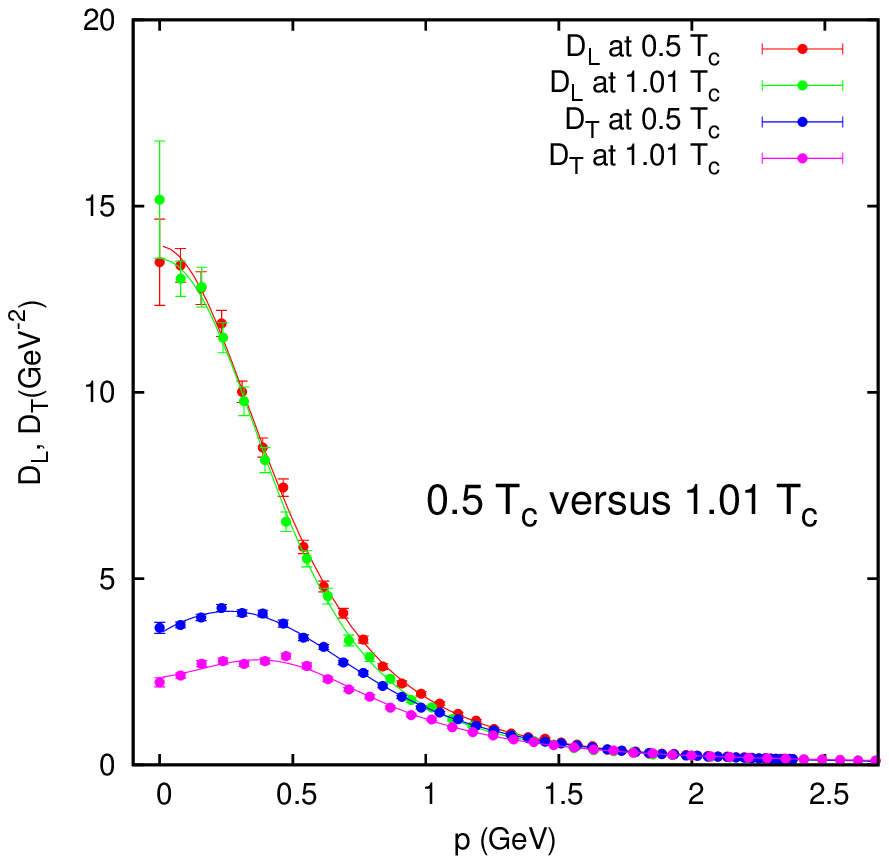}
\caption{Longitudinal and transverse gluon propagators at $T=0$ (top left), 
$T=0.25 T_c$ (top right) and $T=0.5T_c$ (bottom left). Curves for
$T=0.5T_c$ and 1.01 $T_c$ are shown together for comparison on the
bottom right.
Values for $N_s^3 \times N_t$, $\,\beta$, lattice spacing $a$ and spatial
lattice size $L$ (both in fm, in parentheses) are given in the plot
labels, with the exception of the bottom right plot, which is described 
in the text.}
\label{combined}
\end{figure}

We also did runs at low temperatures, namely at $\,T=0$, $\,0.25 T_c$ 
and $0\,.5 T_c$.
We show combined $D_L(p)$ and $D_T(p)$ data for these runs
in Fig.\ \ref{combined}.
We see that $D_L(p)$ increases as the temperature is switched on, 
while $D_T(p)$ decreases slightly, showing a clear turnover point at
around 350 MeV. (Note that the runs at 0.5 $T_c$ on $48^3\times 8$
lattices for $\beta = $ 2.299, 2.301 are equivalent.)
As pointed out before in \cite{Cucchieri:2011ga}, the infrared behavior 
of $D_L(p)$ remains unchanged (within errors) from $0.5 T_c$ to $T_c$.
This can be seen on the bottom right plot of Fig.\ \ref{combined}, 
where we show (for clarity) only lattices with large enough $N_t$ and
the largest physical size. More precisely, we take $\beta=2.299$ on
a $96^3 \times 8$ lattice, corresponding to 0.5 $T_c$, and 
$\beta=2.515$ on a $192^3 \times 8$ lattice, corresponding to 1.01 $T_c$.
We see that, whereas the behavior of $D_T(p)$ is consistent with a 
steady monotonic decrease with temperature, the fact that $D_L(p)$
stays invariant might suggest a flat curve for the infrared-plateau value 
of the longitudinal propagator as a function of temperature below $T_c$.

\begin{figure}
\hspace*{-1.5cm}
\includegraphics[height=7.2truecm]{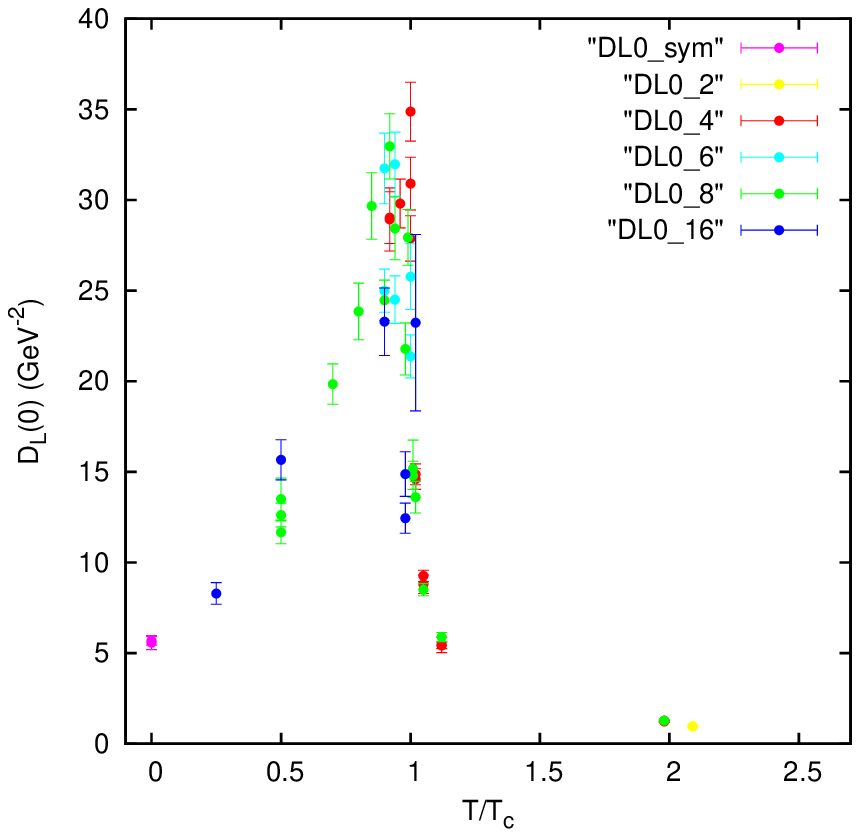}
\hspace*{-2.7cm}
\includegraphics[height=7.2truecm]{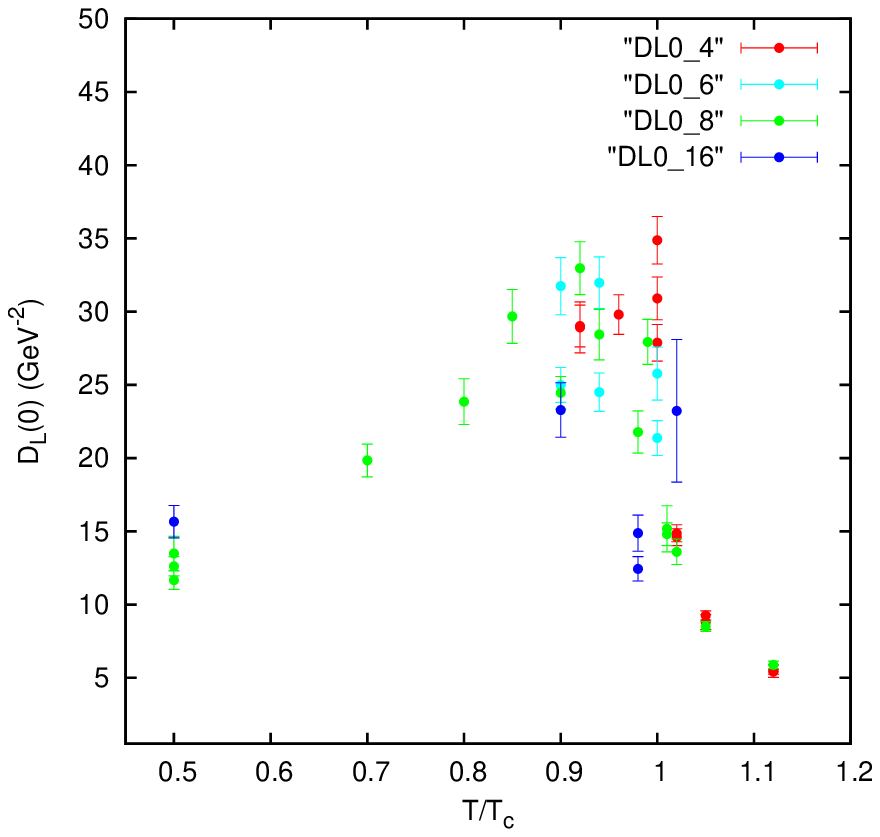}
\caption{Infrared-plateau value for the longitudinal gluon propagator
[estimated by $D_L(0)$] as a function of the temperature for the
full range of $T/T_c$ values (left) and for the region around $T_c$
(right). 
Data points from runs at the same value of $N_t$ are grouped together
and indicated by the label ``DL0\_$N_t$'', where ``sym'' is used to
indicate symmetric lattices (i.e.\ $T=0$).
}
\label{plateau}
\end{figure}

To investigate the issue, we have performed runs at other values ot 
$T\leq T_c$. We have considered several values of $T/T_c$, and studied
the dependence of the infrared-plateau value with $T/T_c$. In
Fig.\ \ref{plateau}, we show data for $D_L(0)$ for all our runs
on the left-hand side, and for the region around $T_c$ on the right.
We group together results from runs using the same value of $N_t$,
and indicate them by the label ``DL0\_$N_t$''.
The data points indicated with ``sym'' correspond to symmetric lattices,
i.e.\ to the zero-temperature case.
Note that results for different $N_s$'s at fixed $N_t$ may not fall on 
top of each other, which gives us an indication of the systematic errors 
discussed above. 
These are especially serious for $N_t=4$ around $T_c$ (red points).
We see that, surprisingly, the maximum value of $D_L(0)$ is not
attained for $T=T_c$ --- as might have appeared to be the case
from the $N_t=4$ lattices only --- and it does not describe a flat 
curve from 0.5 $T_c$ to $T_c$, as could be expected by
looking at the bottom right plot in Fig.\ \ref{combined}. Rather, 
it seems to lie at about 0.9 $T_c$. 
Also, it clearly corresponds to a finite peak, which does not
turn into a divergence as $N_s$ is increased at fixed $N_t$.

\begin{figure}
\hspace*{-1.5cm}
\includegraphics[height=7.2truecm]{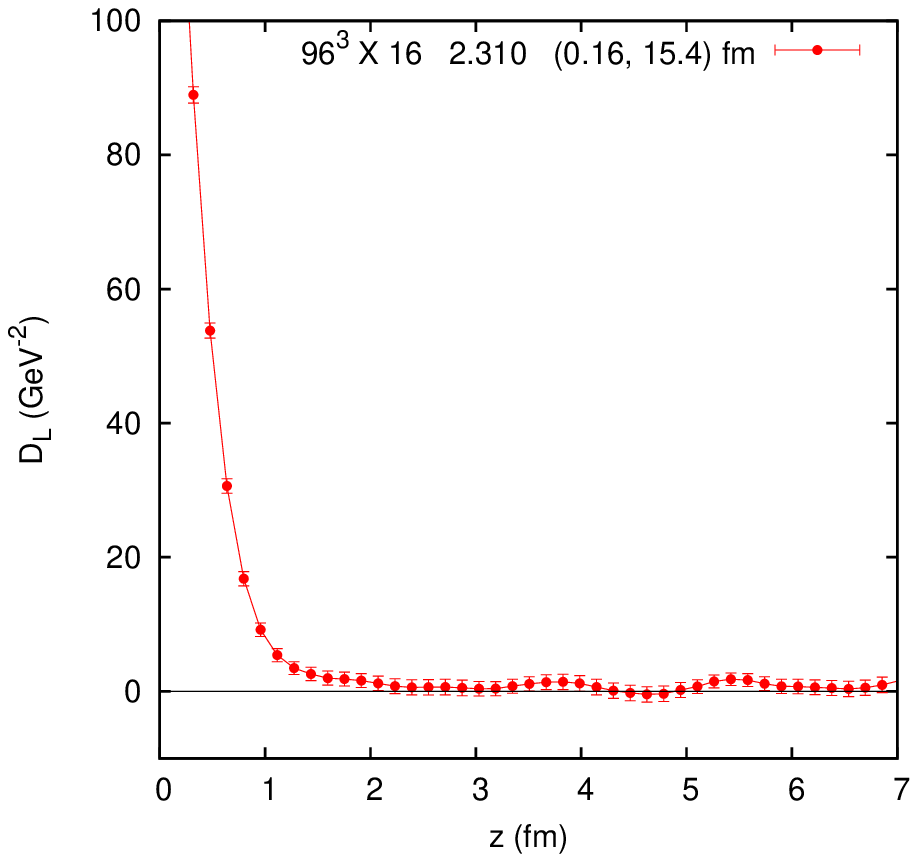}
\hspace*{-2.7cm}
\includegraphics[height=7.2truecm]{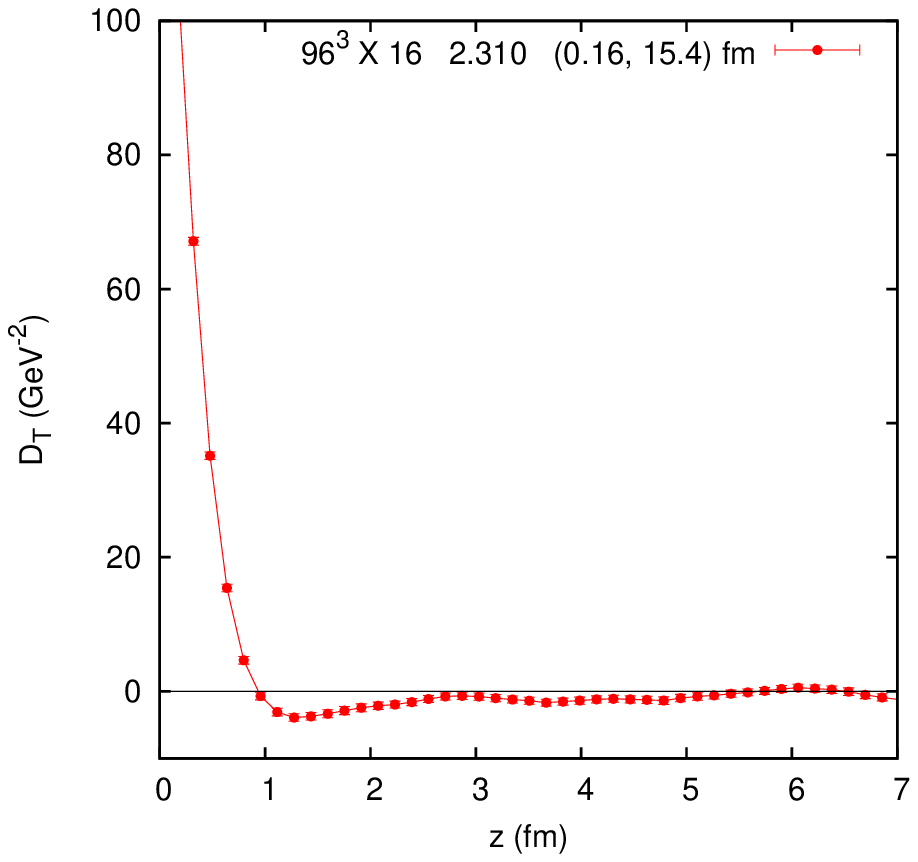}
\caption{Longitudinal (left) and transverse (right) gluon propagator
in real space for $T = 0.25 T_c$.
Values for $\,N_s^3 \times N_t$, $\,\beta$, lattice spacing $a$ and spatial
lattice size $L$ (both in fm, in parentheses) are given in the plot
labels. Note that the solid lines are {\em not} fits.}
\label{zgluon}
\end{figure}

Finally, we also looked at the real-space propagators.
We find clear violation of reflection positivity for the transverse
propagator at all temperatures. For the longitudinal propagator,
positivity violation is observed unequivocally only at zero temperature 
and for a few cases around the critical region, in association with 
the severe systematic errors discussed above. For all other cases,
there is no violation within errors. Also, we always observe an
oscillatory behavior, indicative of a complex-mass pole.
Typical curves for the longitudinal and transverse propagators
in real space are shown (for $T=0.25T_c$) in Fig.\ \ref{zgluon}.


\section{Conclusions}

The transverse gluon propagator $D_T(p)$ shows infrared suppression 
and a turnover in momentum (in agreement with the dimensional-reduction 
picture) at all nonzero temperatures considered. Also, it exhibits
violation of reflection positivity as a function of real-space coordinates
in all cases studied. The longitudinal propagator $D_L(p)$, on the contrary, 
appears to reach a plateau at small momenta, and does not in general
show violation of reflection positivity.
We have obtained good fits of our data to a Gribov-Stingl form, 
with comparable real and imaginary parts of the pole masses, also
in the longitudinal-propagator case. This is in contrast with an
electric screening mass defined by the expression $D_L(0)^{-1/2}$,
which moreover may contain significant finite-size effects.

The data for $D_L(p)$ are subject to sizeable
discretization errors around the critical temperature. In particular,
a severe dependence on the aspect ratio $N_t/N_s$ is seen at 
$\,T\ltapprox T_c$ for the smaller fixed values of $N_t$.
As a result, only lattices with $N_t> 8$ seem to be free from
systematic errors. After these errors are removed, we see an infrared 
value about 50\% smaller than before. As noted in \cite{Cucchieri:2011ga}, 
this might suggest that there is no jump in the infrared value of $D_L(p)$ 
as $T\to T_c\,$ from below, since the resulting infrared behavior
at $T_c$ is essentially the same as at $0.5 T_c$ (see Fig.\ \ref{combined}).
[We note that all previous studies of $D_L(p)$ around $T_c$ had
employed $N_t\leq 4$.] An investigation of the temperature range between
$0.5 T_c$ and $T_c$ shows, nevertheless, a (finite) maximum of the plateau 
value $D_L(0)$ at around 0.9 $T_c$ (see Fig.\ \ref{plateau}).


\section*{Acknowledgements}

The authors thank agencies FAPESP and CNPq for financial support
and the organizers of ``The many faces of QCD'' for a pleasant and
stimulating meeting.
Our simulations were per\-formed on the new CPU/GPU cluster at
IFSC--USP (obtained through a FAPESP grant).

\end{document}